# Programming Technologies for the Development of Web-Based Platform for Digital Psychological Tools

Evgeny Nikulchev[1], Dmitry Ilin[2]
MIREA – Russian Technological
University & Russian Academy
Science, Moscow, Russia

Pavel Kolyasnikov[3]
Vladimir Belov[4]
Russian Academy Science
Moscow, Russia

Ilya Zakharov[5], Sergey Malykh[6]
Psychological Institute of Russian
Academy of Education
Moscow, Russia

*Abstract*—**The choice of the tools and programming technologies for information systems creation is relevant. For every projected system, it is necessary to define a number of criteria for development environment, used libraries and technologies. The paper describes the choice of technological solutions using the example of the developed web-based platform of the Russian Academy of Education. This platform is used to provide information support for the activities of psychologists in their research (including population and longitudinal researches). There are following system features: large scale and significant amount of developing time that needs implementation and ensuring the guaranteed computing reliability of a wide range of digital tools used in psychological research; ensuring functioning in different environments when conducting mass research in schools that have different characteristics of computing resources and communication channels; possibility of services scaling; security and privacy of data; use of technologies and programming tools that would ensure the compatibility and conversion of data with other tools of psychological research processing. Some criteria were introduced for the developed system. These criteria take into account the feature of the functioning and life cycle of the software. A specific example shows the selection of appropriate technological solutions.**

## I. INTRODUCTION

Currently, computer technologies are actively used for data collection in the field of education. In recent years, web-based technologies are going to be widely known for psychological researches. Computers are used not only for questionnaires automation, but also for complex cognitive tests, that contains different graphics. One of the first attempts for cognitive tests automation was undertaken in the TAPAC (Totally Automated Psychological Assessment Console [1]) system that consisted of a console for test subject's answers, tape recoder and projector. Since then, a number of technological resources for psychological testing automation have increased. Internet technologies development radically changed possibility for data collection. First, they make it possible to increase the data set of research. Secondly, web-based technologies reduce the time of data collection and the cost of studies [2]. The development of modern computing technologies provides new opportunities for organizing large-scale population studies of the psychological characteristics of students. The results of these studies can be used for the national standardization of psychometric tools.

In addition, large accumulated data sets can become the basis for machine learning mechanisms and other approaches using artificial intelligence. Accumulation of data from population studies into a single system can allow a breakthrough in the development of systems for automated intellectual analysis of behavior data.

The issue of selecting methodological tools for online and offline research includes several items.

First, any selection presupposes the existence of generally well-defined criteria, on the basis of which a decision can be made to include or not to include techniques in the final toolkit.

Secondly, an important factor in the ongoing research is the separation of them into online and offline methods. Creation of new research tools based on web technologies will allow creating not only complex experimental psychological models, but also increase the statistical power of the received data due to the expansion of research samples. The choice of methods based on this division, on the one hand, imposes certain restrictions on the selection of tools, on the other hand, allows to focus on the strengths and weaknesses of the methods used.

The selection of methodological tools in the field of education and psychology has a number of common for methods of studying the behavior of criteria (requirements for reliability and validity of methods, etc.), and specific parameters (related to the field of application). Common criteria for the selection of tools include generally accepted requirements for the reliability of psychological testing tools. These include:

- Assessment of the quality of testing, which includes primarily an analysis of the knowledge, skills, abilities or individual characteristics that need to be assessed; quality assessment involves the construction of a specific goal and test criteria.

- Validity of the tool (validity and suitability of application of methods and results of research in specific conditions).

- The reliability of the tool (the possibility of obtaining identical results in subjects in different cases).

- The reasonableness of the methodology, the presence of an adequate psychological theory, which underlies the methodology.

- The conformity of the methodological tool to the cultural norms of the estimated population. This is one of the important indicators, which is often underestimated, assumes the account of the so-called "bayes" or hidden context, understood by the testing participant in the answers to the questions of the methods, depending on the cultural norms. This includes the equality of all participants in testing, the absence of harassment on the basis of gender, national, religious grounds, any issues that may somehow hurt feelings or adversely affect a participant.

- The existence of an explicit empirical mechanism for interpreting assessments of the subject, such as population norms, criteria for determining clinical groups and criteria for classifying individuals.

- The quality of administrative, interpretive and technical guidelines. The presence of a clear procedure for the conduct and interpretation of individual techniques, as well as batteries in general. Uniformity in the interpretation of the results obtained and their use.

- The basis for conclusions about broader underlying behaviors and attributes from the pattern of behavior.

- Ease of use of the test material.

Amazon Mechanical Turk platform was the first widely used web-based psychological tool. This platform allowed researchers for a small amount of time to hire individual participants to fill out individual psychological techniques online. MTurk was used mainly by researchers from the United States. It was shown that MTurk samples do not fully correspond to the characteristics of the US population. For example, the population of MTurk was mostly white and female, and also more educated and younger than the US population as a whole. However, the quality of the data obtained remained fairly good [3]. MTurk was successfully used to study the attention [4], creativity, dishonest behavior [5] and sexual attitudes [6]. It has been shown that MTurk can be a valuable tool even for working with a clinical population [7]. To date, the Russian-language Yandex.Tolok platform, similar to the MTurk platform, has been developed. After the success of MTurk, new technologies began to appear, but the approach has changed. While MTurk was an online marketplace with an audience of its own, new tools provided only tools for creating research, which could then be distributed over the Internet, while the researcher himself had to provide a sample. Google Forms service (Google Inc.) was one of the most famous tools of this type. This free software was designed to create web surveys. Every researcher could create his own questionnaire, for which he was given a unique link that he could distribute on the Internet. Google Forms also provided simple analytics for researchers. Its openness and brand Google made it ubiquitous for online surveys, especially among students. However, Google Forms does not have a number of functions that play an important role in psychological research. For example, an analysis of the response time characteristics based on the Google Form Service is not possible. There are also problems with storing the collected data. In this regard, for psychological needs, additional products have been developed. Among the most recognized we can mention Survey Monkey, Qualtrics, LimeSurvey or EnKlikAnketa. Some of these products are paid, or at least shareware (shareware: SurveyMonkey, Qualtrics), some of them are free (EnKlikAnketa). These services can provide many useful functions for researchers. For example, the Slovenian service EnKlikAnketa offers assistance in correcting methodological deficiencies in the development of surveys. It also allows to collect a lot of metadata, such as the site from which the respondent went to fill out the survey, the time spent on the poll, or the characteristics of the browser and operating system of the computer on which the respondent was working.

To date, this type of services has almost replaced traditional research using paper forms. However, they can only be used for questionnaires or knowledge tests (q-type data according to Cattell [8]), whereas studies in the field, for example, of individual differences in cognitive or control functions require a wider functional. In these areas, a class of modern computerized technologies is being developed. Computerized presentation of tests is convenient because of several factors. First, it helps to automatically control the process of presenting tasks, thereby reducing errors related to the human factor. Secondly, computerized tools of this type can record aspects of performance with accuracy not available for other methods. Among such characteristics of the tests, one can single out an estimate of the reaction time, an estimate of the exact spatial position of the cursor on the screen, and others. Various cognitive tests are often organized into batteries that are designed for a comprehensive assessment of the cognitive domain. Widely used cognitive characteristics include general cognitive abilities, working memory [9], spatial or mathematical abilities [10] Most of these batteries have been developed for clinical use NAB [11]. However, they have also been used successfully to study regulatory samples [12] , as well as for research in the field of behavioral genetics [13].

The computerized application of cognitive tests is becoming more accessible to researchers in the field of psychology. First, there are already free software solutions, such as PsychoPy (http://www.psychopy.org/), which allow psychologists to develop their own tests in the absence of advanced programming skills. Some of the applications of this type even contain their own battery tests. For example, there is a programming language for creating tests for psychology (PEBL, http://pebl.sourceforge.net/). It requires more skills to create a specific test than PsychoPy, however, a large set of pre-programmed executables that can measure a wide range of characteristics is freely available to it.

At present, the main limitation of automatic batteries is their possibility to be controlled only autonomously with the help of downloaded and pre-installed software. The psychological community tends not to trust the accuracy of the tests conducted online, due to potential side variables, such as the technical properties of personal computers or respondents' monitors. However, the quantification of technical noise shows that at least for some types of tasks, web experiments can be an acceptable source of data [14]. Thus, the next logical step for computerized cognitive tests is to distribute via the Internet in the same way that it happened with the questionnaires. In accordance with this, the PsychoPy team, for example, recently presented the possibility of launching experiments in a web browser.

Summarizing, it is possible to formulate the main advantages of computerized and web technologies for research in the field of psychology. The main advantages are:

- accessibility for large-scale research;

- the increase in reliability and the potential for generalization of the results obtained;

- lower costs for equipment of premises for experiments;

- the opportunity to avoid all the troubles associated with the use of laboratories: (booking, limited space, the need for expensive specialized equipment, etc.);

- the ability to provide tools around the clock without any time limits;

- the possibility of open research, with a fully voluntary participation, which usually improves the motivation of respondents.

It is also necessary to remember the potential difficulties associated with this type of research. Most of the difficulties can be related to technical control over the comparability of tests conducted on computers with different system characteristics. In addition, the researcher's control over the progress of the test is reduced (the research participant can perform tasks alone without additional supervision).

At present, many tools are used for software development. These tools differ in their functionalities and programming convenience, as well as they are not without disadvantages, that often appear only in development stage, when system extension or modules integrating. Therefore, it is an important task to choose toolset and programming technologies in the planning stage. This choice should satisfy the software requirements and programming process. In this case, it is necessary to consider the parameters of the technologies [15], the guaranteed quality of data processing [16], reliability with extension [17] etc.

The aim of the paper is to describe the choice process of the technological solutions on the example of the being developed digital web-based platform. The platform provides information support for psychologists' activities in conducting research (including population and longitudinal researches) [12, 18].

Software architecture is not only a structural basis for system components and their connections describing, but also determines the approaches to development and environment. The architecture description should include answers to the questions that arose during the system designing.Described service is web-based platform that includes server and client sides. Therefore, one of the main tasks is to choose the programming language and technologies that are most suitable for components development [19].The platform should work in most browsers, including mobile devices, without installing any plug-ins or extensions. Therefore, it is necessary to choose the solutions that will not make any certain restrictions or need installing additional plug-ins and libraries.The requirements for server-side components are less restrictive. However, it is necessary to take into account the features of the technological solutions and the complexity of the result software.

Therefore, there are following system features:

- significant amount of developing time that needs implementation and ensuring the guaranteed computing reliability of a wide range of digital tools used in psychological research;

- ensuring functioning in different environments when conducting mass research in schools that have different characteristics of computing resources and communication channels;

- possibility of services scaling;

- security and privacy of data;

- use of technologies and programming tools that would ensure the compatibility and conversion of data with other tools of psychological research processing.

To achieve effective solution a number of techniques were used. The primary stage included architecture requirement analysis. The aim of this stage is to identify the main use case, functional and non-functional requirements for the web-based platform [20].

Based on the received information, architecture synthesis was carried out. The reason is to determine a set of coupled components of the system, their connections, the most effective ways of data exchanging.To choose concrete programming language and technologies usable for web-based platform development their study and comparing were carried out. It was completed in the context of the formed architecture, existing requirements and constraints. As for programming languages for browser applications, the ability of application delivering without the need for installation additional software was evaluated. Frameworks are considered for their active application in projects, their community and relevance of the task. It is worth to note that the direct comparison of frameworks will not give concrete result. However, a number of these frameworks are more suitable due to better scalability, less costs for study and more ready-made modules.

The paper includes following ensuing sections. The IIand III parts describe the features of the being developed system architecture and its main components. In the IV part, criteria are introduced for the client and server sides of the application. The section V describes the choosing of the appropriate technological solutions.

## II. PLATFORM ARCHITECTURE DESCRIPTION

For the formation of an adequate architectural solution, a number of methods have been applied. The initial phase included the architectural requirement analysis, in order to identify the main uses, functional and non-functional requirements for the platform [20]. In addition, to clarify the requirements, unstructured interviews of the pedagogical staff were used to reveal the degree of variation in the technical characteristics of software and hardware in Russian schools.

From the perspective of the end-user, the project will consist of two main components: the researcher's private account (Fig. 1) and offline applications for offline experiments (Fig. 2).

Given the fact that the number of users will grow over time, the web service should be scalable horizontally (Fig. 3). Every API node should include a multi-level architecture. In combination with Object-DocumentMapper (ODM), it will provide more flexibility than monolithic architecture.

Experimental and intensive data algorithms for demographic research should be separated from the main service, as well as from administrative functions (Fig. 4). From a security perspective, the administration panel can be used as a separate service on the intranet.

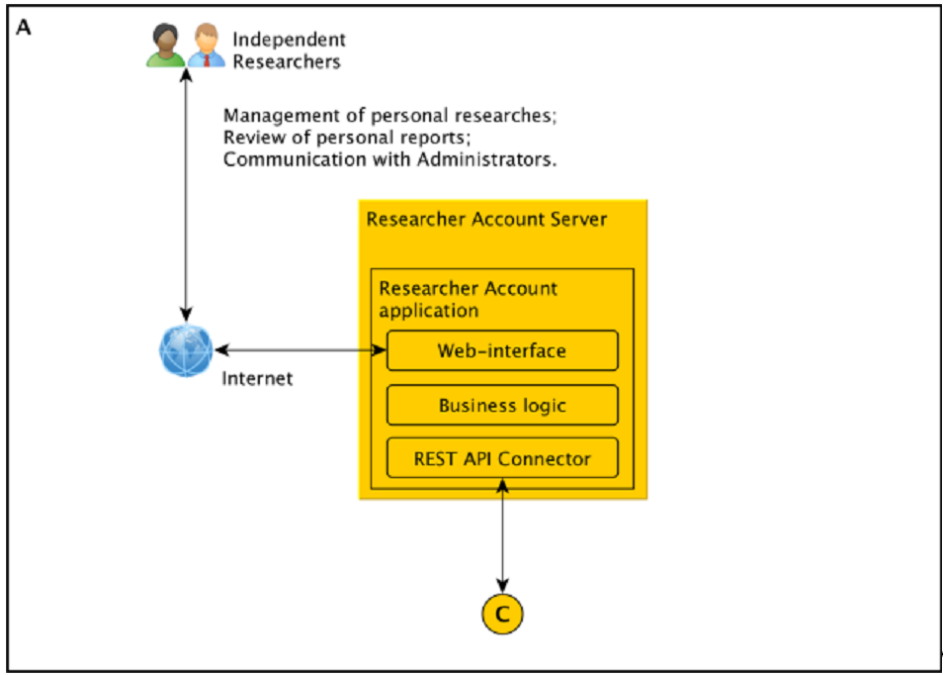

Fig. 1. An Independent Researcher Tool.

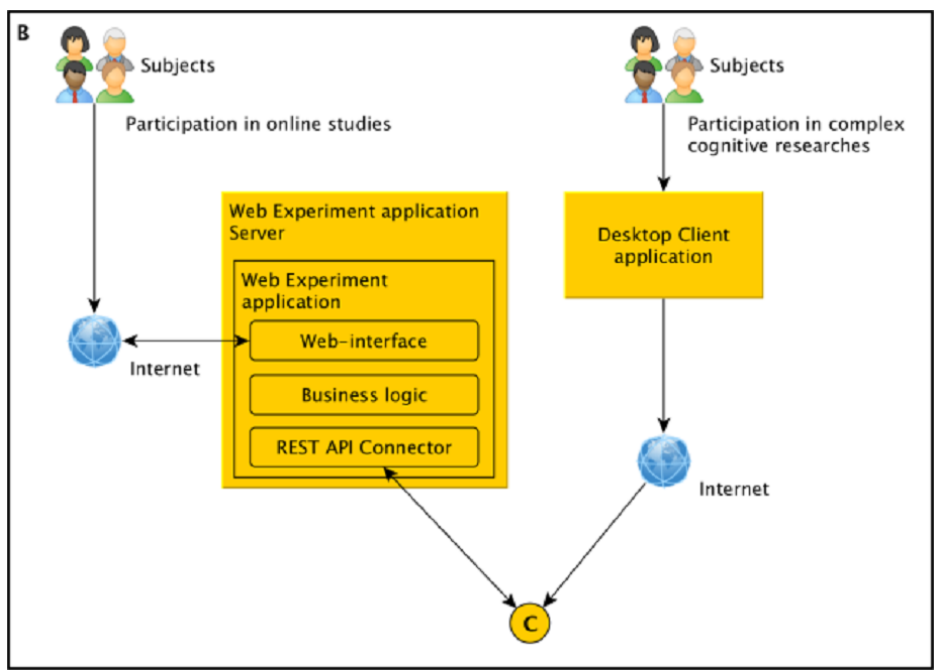

Fig. 2. Tools for Online and Offline Experiments.

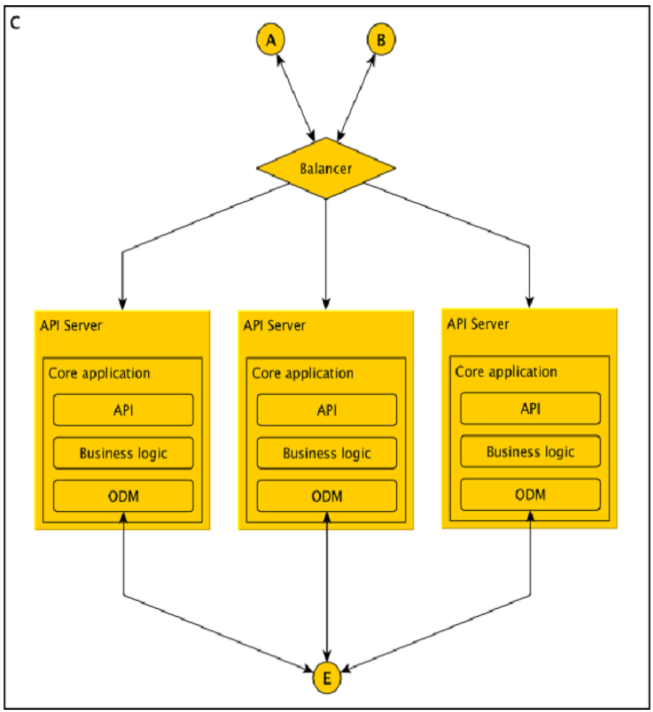

Fig. 3.    Scalable API Server for Unified Data Access and Management.

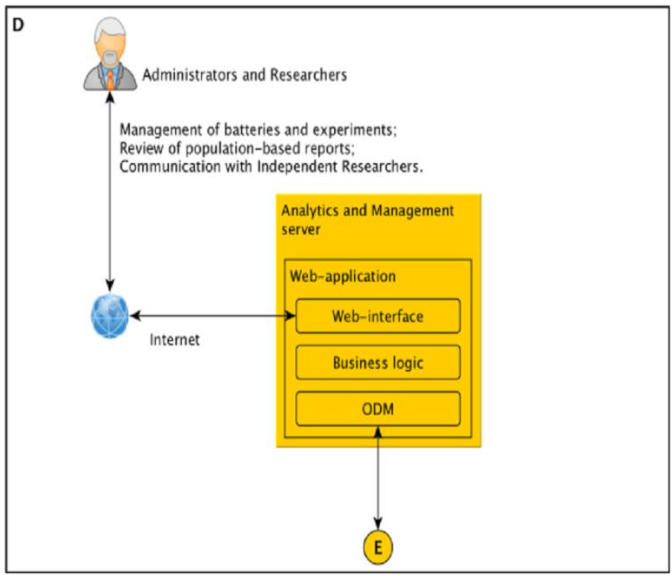

Fig. 4.    Isolated Analytical Tools for Population Studies.

In addition to scaling the algorithmic part, data storage should also be scalable [21, 22]. The best approach for the project is the combination of sharding and replication (Fig. 5).Sharding allows to distribute data between different physical servers (shards) based on the value of some key, so that the entities are grouped into data set for this key.Replication allows to copy data between several servers, among which one server (master) for data saving, and others (slave) - for reading.Thus, sharding can provide a system with high I/O performance, while replication can help to ensure the availability of the service.

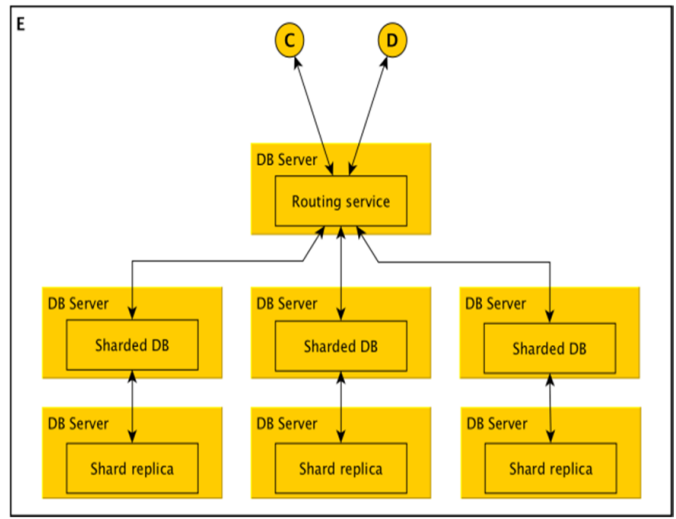

Fig. 5.    Horizontal Scaling of Storage by Means of Sharding and Replication.

## III.  Description of Platform Components

The architecture of the being developed platform for psychological research was chosen to be multi-component, which provides more flexibility than monolithic.

Monolithic architectures have a number of disadvantages:

- the larger the system, the more difficult it is to maintain it and make changes;

- with a large system, changing a small portion of the code can cause errors in the entire system;

- after each code change, it is necessary to test the entire system for errors.

Unlike monolithic, the use of multi-component architecture gives the following advantages:

- writing and maintaining smaller parts is easier than one large system;

- it is easier to distribute the developers to write a specific part of the system;

- the system can be heterogeneous, because for each component it is possible to use own languages and technologies, depending on the task;

- easier to upgrade because only the required component is affected;

- the system becomes more fault-tolerant, since in the event of failure of one and the components, others may still be working.

Thus, the choice in favor of a multi-component architecture is justified in view of a number of advantages over the monolithic and the most suitable taking into account the requirements for the developed platform.

Fig. 6 shows the scheme of the platform architecture for psychological research. The architecture is divided into separate components that can work independently and communicate among themselves using the REST API.

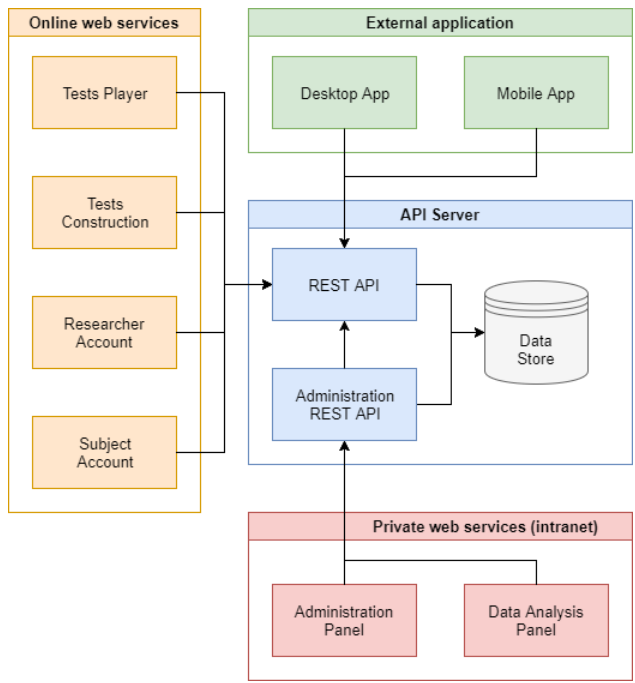

Fig. 6.  Schematic of a Multi-Component Platform Architecture.

**«APIServer»** is the main core of the system, which is a RESTAPI server and is responsible for working with the data store, as well as for performing various service functions.

**«Onlinewebservices»** are components that should be accessible from the Internet. They are the main online part of the platform being developed, among which are the online test player, the online test designer, the researcher's personal area and the personal area of the examinee.

**«Externalapplication»** is separate application, such as desktop and mobile. Unlike the online version of the test player, the feature of the applications is that the process of passing these tests should work without connecting to the Internet. In this regard, the data for the tests should be loaded in advance, and after passing the tests the subject is uploaded back to the server.

**«Privatewebservices (intranet)»** are separate services, including the Platform Administration Panel and the Population Analysis Toolbar. The feature of these services is their need to be isolated from direct access from the Internet for a greater security guarantee. It is also worth noting that these services communicate with their own RESTAPI, which includes administrative methods, which should not be accessible from the Internet.

To carry out psychological research, many tests are used, which can include necessary materials, for example, images or files.

Thus, for storage and transmission of tests, a batch approach is relevant, which will allow storing and transferring data in one file. The use of the package is also important for transferring data to the client part of the application, where it cannot always be guaranteed access to the Internet. It includes both an online application and a desktop application that will be used in schools, as well as an application for mobile devices. It is possible to draw an analogy of the test suite with data for city maps, which are downloaded as a package and then unpacked onto the device.

Packages will be used both for sending psychological tests to the application and for obtaining test results and sending them to the server.

Using a package for data transferring to online applications in the client browser is justified by following reasons:

- one request to the server is used, instead of several, which minimizes the load on creating an HTTP connection;

- the client side does not depend on the server at the time of passing the test;

- in the case of Internet disconnection, the user can complete his research;

- logging time on the client side minimizes the errors of the record;

- it is easier to track the degree of workload of the package and to inform the user about it.

Fig. 7 shows the structure of the package for storing the tests. It shows that the package is complete, includes information and description of this package, as well as the tests themselves. There should be at least one test in the package. The test includes data on its description, as well as images and files, if they are needed.

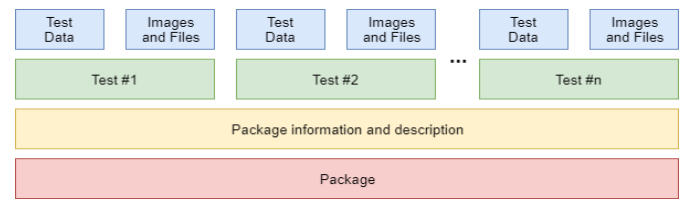

Fig. 7.  Diagram of the Structure of the Package for Storing and Transferring Tests.

The test should be described using a special JSON Schema standard, the structure of which is approved in advance. Based on this structure, the test will be validated.

Using JSON Schema avoids a number of problems and has the following advantages:

- no need to manually check the contents of the documents;

- no need to create own validators with a variety of configurations and takes care of the support of these decisions;

- due to a single standard, the process of integration and support of validation in various components of the platform is simplified, such as online test player and desktop applications;

- the change in the scheme does not require the replacement of the validator code;

- it is possible to describe the psychological test manually without the help of additional tools, which will be a plus at an early stage of development, when the test designer is not yet available;

- there are a large number of implementations for various programming languages and platforms.

Based on the requirements to the platform for digital psychological research, a list of virtual machines required for their deployment on the server was generated (Table I).

TABLE I. VIRTUAL MACHINES

| Purpose of the VM | Number of VM (pcs) | OS Version | RAM (GB) | CPU (pcs) |
|---|---|---|---|---|
| API server | 2 | CentOS 7.3 | 8 | 4 |
| Personal cabinet of a psychologist | 1 | CentOS 7.3 | 2 | 1 |
| Online survey system | 1 | CentOS 7.3 | 2 | 1 |
| Internet portal | 1 | CentOS 7.3 | 2 | 1 |
| Intranet administration system | 1 | CentOS 7.3 | 2 | 1 |
| Intranet-system of analysis of population data | 1 | CentOS 7.3 | 2 | 1 |
| MongoDB request router | 1 | CentOS 7.3 | 1 | 2 |
| MongoDB configuration server | 1 | CentOS 7.3 | 1 | 2 |
| MongoDB shards | 3 | CentOS 7.3 | 2 | 2 |
| Stage-server | 1 | CentOS 7.3 | 8 | 4 |
| Jenkins Continuous Integration System | 1 | CentOS 7.3 | 4 | 2 |
| DNS, DHCP | 1 | CentOS 7.3 or win 2012r2 if there is a subscription | 4 | 2 |
| Backup virtual machines | 6 | CentOS 7.3 or win 2012r2 if there is a subscription | | |

## IV. THE CHOICE OF TECHNOLOGICAL SOLUTIONS FOR THE DEVELOPMENT OF THE CLIENT PART

Based on the information received, architecture synthesis was carried out to determine a set of loosely coupled components of the system, their connections, and the most effective ways of exchanging data.

To choose the programming languages and technologies solutions for the platform development, research and comparison were conducted. It was carried out in the context of the formed architecture, requirements and constraints.

For the programming languages and technologies analysis, reports and materials of services such as StackOverflow and GitHub, which are the most authoritative in the software development environment, were studied.

With regard to programming languages for browser applications, application delivery capabilities are evaluated without the need for additional software.

Frameworks are considered for their active application in projects, the size of the community of developers, the relevance of the task and the time on the market.

It should be noted that direct comparison of frameworks for development will not yield results, since each of them will allow to reach the final result. Nevertheless, a number of them should be considered more suitable due to better scalability, less training costs and more ready-made modules.

As a result of the consideration of JavaScript, Java applets and Adobe Flash platform from the point of view of applicability for code execution in the browser, it was found out that only JavaScript can be considered applicable.

This is due not only to the fact that JavaScript is used in many areas (client browsers, server part, mobile platforms), as well as the desktop applications. Java applet technology, like Adobe Flash technology, requires the installation of additional components in the user's system. Moreover, depending on the operating system and the browser, the installation and configuration process may vary.

Due to the great variability of the hardware and software in schools, the use of these two technologies is not advisable, since this can complicate the process of conducting mass research. It should be noted that in browsers on mobile devices Adobe Flash and Java applets are not supported. It's also worth noting that Adobe Flash rejects HTML5, which can work with multimedia (video and audio).

Thus, the choice in favor of the JavaScript language for the development of the client part becomes obvious and there is no alternative solution under the given conditions at the current moment. JavaScript is supported by all common browsers and is included in them by default.

Developing large Single-Page Applications (SPA) based on pure JavaScript on the client side is a difficult and inefficient process, so it is needed to use frameworks that define the application structure and have a basic set of components [23-25].

It should be noted that almost all modern frameworks have similar functionality and are able to solve the task. Thus, the choice should first of all be based not on the functional of the framework, but on the requirements and objectives within the framework of a particular project.

The most famous frameworks were selected for consideration, including Backbone.js, AngularJS 1, AngularJS 2, React, Ember.js, Vue.js and Polymer. Table II shows the advantages and disadvantages of these frameworks, taking into account the applicability to the developed platform.

Backbone.js is ill-suited for developing large projects, as there are no necessary components for implementing complex functionality. Thus, according to the authors of the article, the use of this framework is inexpedient in view of the fact that it does not have sufficient functionality, and there are also alternative solutions.

TABLE II.    ADVANTAGES AND DISADVANTAGES OF FRONT END FRAMEWORKS

| Framework | Advantages | Disadvantages |
|---|---|---|
| Backbone.js [26] | Compact Simple structure Steep learning curve Rich documentation Supports REST | Does not support two-way data binding. Requires additional components to implement complex functionality. Bad for large projects |
| AngularJS 1 [27] | High popularity Steep learning curve Rich documentation Great community Many ready-made solutions It is part of the MEAN stack (MongoDB, Express.JS, AngularJS, NodeJS) Supports REST High speed development Supports two-way data binding | It is believed to be outdated, since there is an AngularJS 2 Not compatible with AngularJS 1. The speed of work decreases with a sufficiently large amount of data |
| AngularJS 2 [28] | Rich documentation Great community Has a large number of functions Supports REST There are Angular Universal for solving problems of search engine optimization (rendering of pages on the server) Supports two-way data binding | Uses TypeScript to compile in JavaScript. Less steep learning curve compared to AngularJS 1. It is necessary to take many actions to solve even a small functional |
| React [29] | Compact High performance Good documentation Suitable for large and complex projects with a high degree of load | Requires additional implementation on the server for working with data (for example, Flux or Redux). Not supported by REST. Not compatible with libraries that modify the DOM. Less steep learning curve. Complex approach to development, unusual for beginners |
| Ember.js [30,31] | Rich documentation Large ecosystem Suitable for complex and large applications Supports REST Supports two-way data binding | It is considered to be monolithic in comparison with other frameworks. There is no reuse of components at the controller level. Less steep learning curve. Heavy structure. Too big for small projects |
| Vue.js [29] | Very rapidly growing popularity Steep learning curve Few dependencies Good performance Rich documentation Good ecosystem Supports two-way data binding | A fairly young framework. Developed mainly by one person. Not many projects were done. No "out of the box" REST support (there is an Axios library) |
| Polymer [32] | New and promising technology Web Components High speed of work | Too young solution. Great risks when using. Few ready solutions and examples. Less steep learning curve |

Polymer is a library that is based on a fairly new technology Web Components technology. The W3C specification for this technology is not yet complete. There may be any problems with browser support, problems in stability of work, and also a large threshold of entry for developers. In this regard, the use of this framework was decided to be abandoned due to possible risks.

React, unlike others, is a library and does not allow to create a web application, since it is designed to create a View part and should work with data on the server, for example, in conjunction with Flux or Redux. Therefore, React is difficult to understand, has an unusual structure, which complicates the understanding of the application as a whole, and also has a large entry threshold for novice developers. According to the authors, React is more difficult to make a quick prototype and support the solution than on another framework.

AngularJS 1, AngularJS 2, Ember.js and Vue.js have two-way data binding, the ability to build large systems, good documentation and community. The choice will be made between these frameworks.

Ember.js has a complex project structure and a large entry threshold for novice developers, and in case of going beyond the standard use is cumbersome and not flexible. In addition, the framework is less popular than AngularJS and Vue.js.

Vue.js version 2 is currently the fastest growing popular framework, it took the best solutions from Ember.js, React and AngularJS, and also has good performance. Another important factor is that Vue.js does not support REST and requires an additional Axios library for this. In addition, the framework is young and is developed mainly by one person, so its use can lead to greater risks.

As a result, the most appropriate for developing a platform for psychological research is AngularJS 1 and AngularJS 2. AngularJS 1 is a fairly simple framework for mastering and understanding, has a low entry threshold with a rich set of functions. AngularJS 2 is a parallel project with AngularJS 1 and is developed separately. AngularJS 2 greatly complicated, for writing the simplest application requires much more action. In addition, it is written in TypeScript, which will require additional knowledge from the developers.

Taking into account what was written above, as well as the fact that the developed platform for psychological research has a limitation in resources and involves novice developers, and also the most appropriate solution for the current moment, according to the authors of the article, is AngularJS 1. In addition, AngularJS 1 has more popularity than other frameworks, according to GitHub and patent analysis [33].

## V.    SELECTION OF TECHNOLOGICAL SOLUTIONS FOR THE DEVELOPMENT OF THE SERVER PART

The development of the server part of the platform allows to choose from a fairly wide range of technologies, in comparison with the client part. This is primarily due to the fact that server technologies depend on the preferences of developers, equipment and requirements for the project, while client technologies are severely limited. The choice of technological solutions for the development of server components is better to start not with programming languages, but with consideration of frameworks because they set the basic structure for the development of the application, as it was written above. Table III presents the features, advantages and disadvantages of the most suitable frameworks for the development of the server part of the platform.

TABLE III.    ADVANTAGES AND DISADVANTAGES OF SERVER-SIDE FRAMEWORKS

| Framework | Language used | Advantages | Disadvantages |
|---|---|---|---|
| Laravel, Symfony [34] | PHP | Steep learning curve A large number of PHP developers | Blocking IO calls PHP interpreter has low performance No paid support |
| Django [35] | Python | Steep learning curve Generating the administration panel for relational databases | Blocking IO calls Does not support NoSQL solutions out of the box |
| Ruby on Rails [36] | Ruby | Steep learning curve | In the development community, there are references to scaling problems under increasing load. Blocking IO calls |
| Express.js [33] | JavaScript (Node.js) | Not blocking by default (asynchronous) Steep learning curve | Long-term support of the project has difficulties (complexity of refactoring). Development in large groups can be difficult |
| Loopback [37] | JavaScript (Node.js) | Not blocking by default (asynchronous) Generating the Preview Panel and Working with the REST API Declarative approach to the generation of the REST API | The generated API does not contain methods for mass update of related entities |
| Play [38] | Scala / Java | Not blocking by default (asynchronous) Well scaled even with blocking code Strict typing simplifies refactoring | Slow compilation New versions of the framework require improvements in the final software |
| Vaadin [39] | Java | Contains the library of pre-made UI elements Frontend code is generated based on the server Strict typing simplifies refactoring | Blocking by default Slow compilation High threshold of occurrence The development of new UI elements is time-consuming There is no full control over the Frontend code |
| ASP .NET MVC [40] | C# | Strict typing simplifies refactoring | Lock-in to the Windows platform Need to purchase Windows Server licenses for deployment |

Since it was determined that a high degree of project scalability is required, attention should be paid to non-blocking I/O frameworks. In this regard, it is necessary to exclude Laravel, Symfony, Django and Ruby on Rails from consideration. Also, due to the complexities of implementing non-blocking I/O and custom interfaces, the Vaadin framework is not suitable for the project.

ASP .NET MVC imposes additional restrictions on the infrastructure in the absence of significant advantages, so the framework should be excluded from further consideration. Thus, the main choice will be made between the Express.js, Loopback and Play frameworks.

An important factor is the programming language on which the framework is written. Express.js and Loopback are written in Node.js (JavaScript), while Play is developed on Java. In the case of JavaScript, a single syntax will be used for both the client and server parts. This will increase the effectiveness of the development of the platform, since the developer will need to know not two, but only one programming language, which is an advantage in the conditions of a small number of developers. In addition, it allows to combine part of the learning process and to reduce the overall threshold of entry, which will affect the time of training of new professionals who will participate in the development of the platform. Also, JavaScript is the most popular language in the world according to the statistics of such large services as GitHub and StackOverflow. In this regard, according to the article authors opinion, it is more expedient to use the Express.js and Loopback frameworks than Play Framework.

Of the remaining two frameworks, the choice in favor of Loopback is more appropriate for a number of reasons:

- Loopback offers a number of patterns that will help maintain the proper level of support for the code base as it increases;

- The framework is based on Express.js, which will enable all its functional components;

- Loopback offers functionality for simplified API generation, which greatly reduces the amount of labor involved in development.

All of the above mentioned, according to the authors of the article, is more significant than the steepest learning curve. Thus, the choice is stopped on the Loopback framework.

## VI. DISCUSSION

With the use of previously described technologies the following key components of the platform were developed: API Server, Researcher Account, Test Player. Fig. 8 shows the page example of research project list. The example of test visualization for research participant is showed in Fig. 9.

The platform prototype is being tested in trial production. At the moment, the service functionality is partially implemented, but it is already possible to collect data using tests such as Dark Triad, Big Five and a number of techniques for assessing spatial thinking.

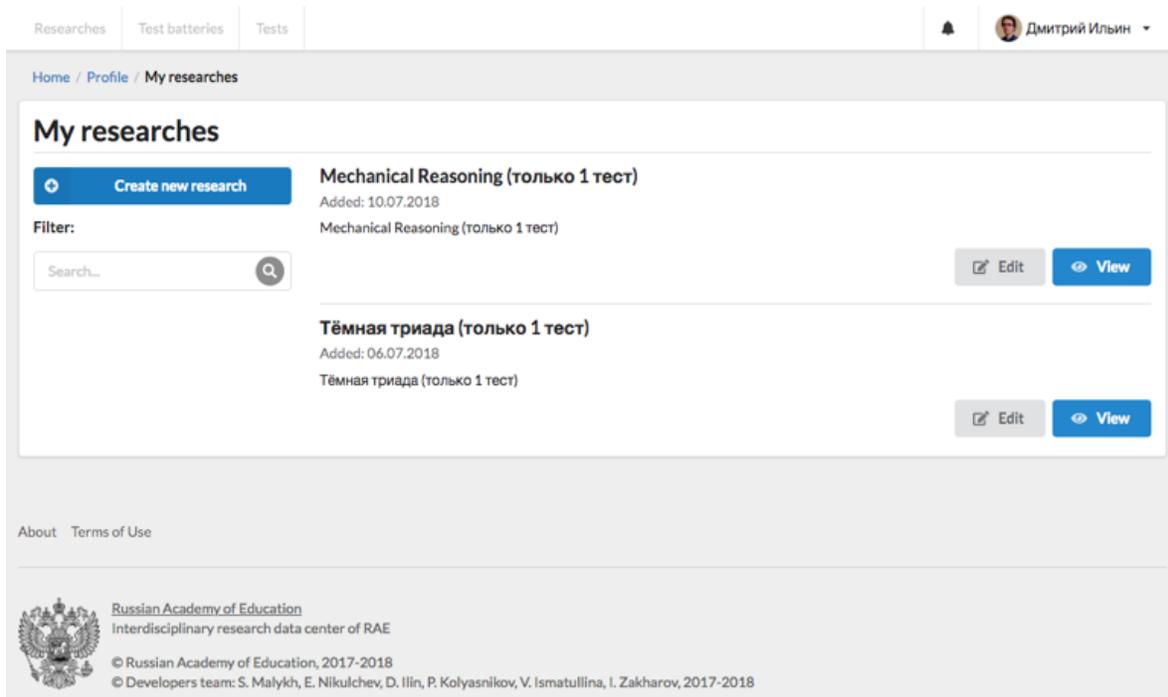

Fig. 8. Researcher Account Page.

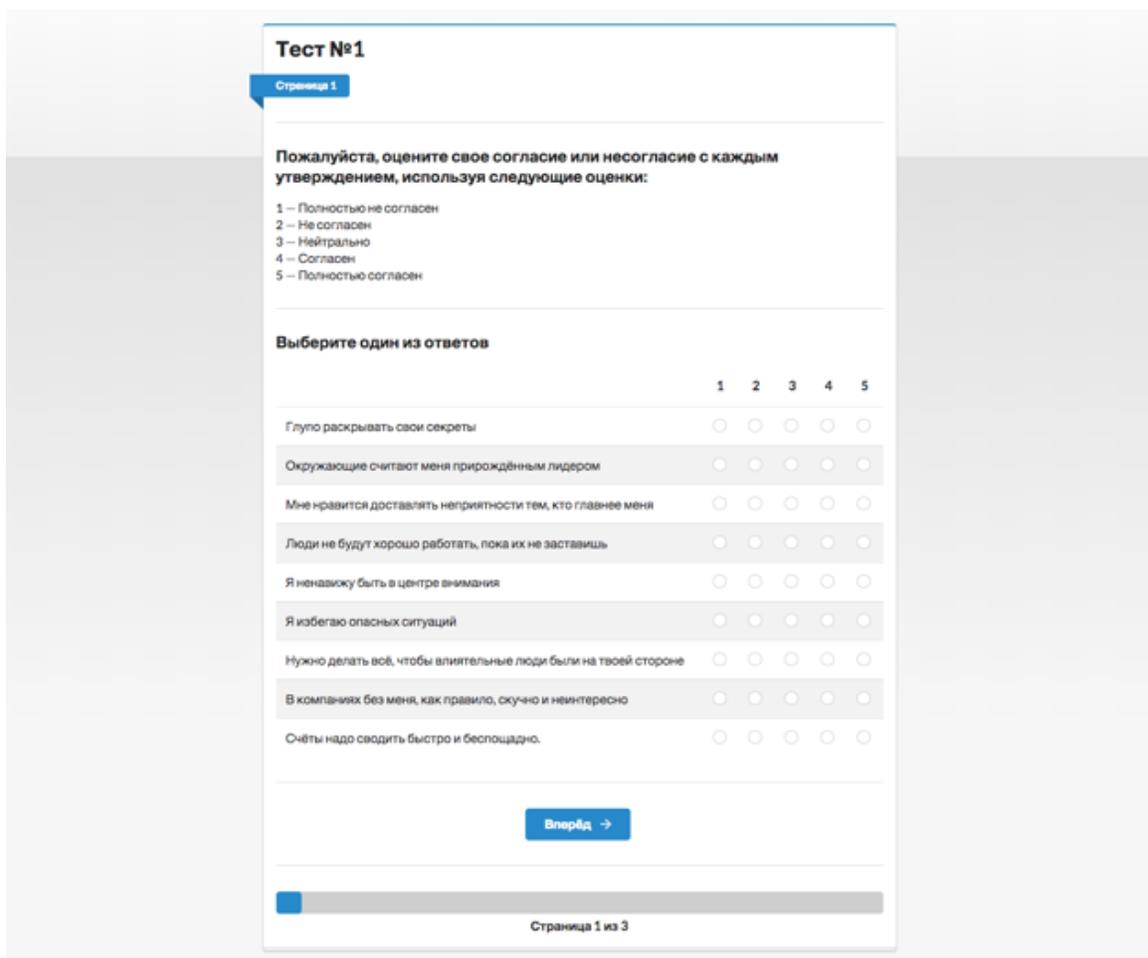

Fig. 9. Pychological Dark Triad Test Visualization.

## VII. CONCLUSION

The guaranteed quality of disturbed system functioning, their effective and success work over a number of years, the scaling ability and connection to different platform are laid at the system designing stage. The chaotic development that was popular ten years ago belonged to the past. There ara following reasons: often to the process of developing completion and software launching used technologies have become outdated. New approaches require careful and comprehensive documenting of the design and implementation process. The following concepts are widely used: ecosystem of programming languages, automated systems and development tools that allow designing software that is able to ensure the quality and reliability of the tasks.

The paper described a concrete example of the development process of the digital web-based platform for psychological research. The choosing of technologies for client and server part was showed. Currently, the system kern is developed and being tested in one of the education institution for software support for psychological research. Developed system showed an adequate choice of software technologies. The system successfully operates in test mode.

In recent years, significant changes in the approaches to the study of problems of education have occurred in the world science. Rapidly developing high-tech methods of studying a person expand the possibilities for studying the mental development and learning of students. Research in this field becomes interdisciplinary and actively assimilates the approaches and methods of the whole series of other sciences. The use of new technologies of human study forms a huge amount of data even on small samples. Providing the conditions for training requires taking into account the age and individual psychological features of modern students. To solve this problem, population studies are needed, which are the basis for determining contemporary age norms of the student 's mental development and national standardization of psychological diagnostic methods.

Significant individual differences exist in activities related to various forms of education (primarily, schooling). These differences are associated with both general cognitive abilities and private cognitive functions: motivation, emotions in response to the learning process, school and family environment. This determines the importance of researching individual differences for all fundamental learning problems. Each person has a unique genetic profile, which in turn forms individual psychological characteristics. The environment adapts to genes with the participation of parents, school and students themselves.

The complex nature of the interaction of genetic and environmental factors of individual differences at the psychological and psychophysiological levels requires the selection of adequate tools for conducting research.

The concept of the architecture of the web-based platform is formulated. A scalable multicomponent architecture is proposed. Scalability of the data warehouse is provided by technologies of sharding and replication. A list of virtual machines for deployment on the server has been generated.

Web-based platform is divided into server part (REST API server and data warehouse), public part (online test player, online test designer, researcher's personal profile and personal profile), private part (administration panel and data analysis panel) and external applications (desktop and mobile). In terms of security, the private part is used as a separate service on the intranet and has its own REST API. To implement the architectural aspects, the most appropriate technologies were chosen in the given task: JavaScript language that will be used to implement most of the software components, AngularJS framework for the client part of the online application, Loopback (Node.js) framework for implementing the API server that provides a single access point for all other components of the platform.The result of the techniques applying is implemented programmatically in the prototype platform, which is being tested in production trial.

### ACKNOWLEDGMENT

The work was financed by the Ministry of Education and Science of Russia, project 25.13253.2018 / 12.1 "Development of the technological concept of the Data Center for Interdisciplinary Research in Education".